\documentclass{webofc}
\usepackage[varg]{txfonts}   % Web of Conferences font
\woctitle{?????????}
\begin{document}
\title{Signals embedded in the radial velocity noise}

\author{Mikko Tuomi\inst{1,2}\fnsep\thanks{\email{mikko.tuomi@utu.fi}}}

\institute{University of Hertfordshire, Centre for Astrophysics Research, Science and Technology Research Institute, College Lane, AL10 9AB, Hatfield, UK
\and University of Turku, Tuorla Observatory, Department of Physics and Astronomy, V\"ais\"al\"antie 20, FI-21500, Piikki\"o, Finland}

\abstract{%
Bayesian data analysis techniques, together with suitable statistical models, can be used to obtain much more information from noisy data than the traditional frequentist methods. For instance, when searching for periodic signals in noisy data, the Bayesian techniques can be used to define exact detection criteria for low-amplitude signals -- the most interesting signals that might correspond to habitable planets. We present an overview of Bayesian techniques and present detailed analyses of the HARPS-TERRA velocities of HD 40307, a nearby star observed to host a candidate habitable planet, to demonstrate in practice the applicability of Bayes' rule to astronomical data.
}
\maketitle
\section{Introduction}\label{sec:intro}

The Bayes' rule of conditional probabilities has a long and controversial history in astronomy and has been used widely in some parts of the astronomical community, such as among cosmologists \cite{trotta2008}, but has only recently been applied to transit or radial velocity planet search data \cite{parviainen2012,tuomi2012,tuomi2013a,tuomi2013b}. The principles of Bayesian inference are remarkably simple but they still enable one to perform data analyses to be with little assumptions that might be limiting the quality of the inferred results when using more traditional \emph{frequentist} data analysis methods.

The Bayes' rule can be written simply as
\begin{equation}\label{bayes}
  \pi(\theta | m) = \frac{l(m | \theta) \pi(\theta)}{P(m)}, P(m) > 0 ,
\end{equation}
where random variable $m$ denotes measurements and random variable $\theta$ represents all the unknowns that are of principal interest to the researcher, i.e. the model parameters. In the above equations, $\pi(\theta | m)$ is the posterior density of the parameter given the measurements; $\pi(\theta)$ is the prior density, the information on $\theta$ before the measurement $m$ was made; $l(m | \theta)$ is the likelihood function of the measurement, i.e. the statistical model or likelihood model; and $P(m)$ is a constant, usually called the marginal integral, that makes sure that the posterior is a proper probability density in the sense that the product of the likelihood and prior integrates to unity over the parameter space $\Theta$. Thus, the marginal integral can be written as
\begin{equation}\label{marginal_integral}
  P(m) = \int_{\theta \in \Theta} l(m | \theta) \pi(\theta) d \theta .
\end{equation}

The integral in Eq. (\ref{marginal_integral}) is a key feature in the Bayesian model comparison framework. However, we do not discuss model selection further in this article and simply refer to some recent papers discussing the caveats of estimating $P(m)$, and methods of overcoming them and using it in practice \cite{kass1995,chib2001,clyde2007,tuomi2012b}

The Bayes' rule works the same way regardless of the number of measurements (or sets of measurements) available. For independent measurements $m_{i}, i=1, ..., N$, we can generalise Eq. (\ref{bayes}) and write
\begin{equation}\label{bayes2}
  \pi(\theta | m_{1}, ..., m_{N}) = \frac{l(m_{1}, ..., m_{N} | \theta) \pi(\theta)}{P(m_{1}, ..., m_{N})} = \frac{\pi(\theta) \prod_{i=1}^{N} l_{i}(m_{i} | \theta)}{P(m_{1}, ..., m_{N})} ,
\end{equation}
where the last equality holds only if the measurements are independent and thus $l(m_{1}, ..., m_{N} | \theta) = \prod_{i}l(m_{i} | \theta)$ holds.

The expression in Eq. (\ref{bayes2}) holds with remarkable generality in practice and its applicability is practically without limits. For instance, no assumptions are needed on the nature of the likelihood function, sometimes called the likelihood model, as long as it can be expressed mathematically. Furthermore, the measurements can be anything: from different sources such as radial velocity, transit, or astrometric surveys in the context of planet searches; time-series, individual numbers, matrices such as digital images, etc.; integers, real numbers, or even boolean operators. We also note that no assumptions are required about the nature of the probability densities of model parameters $\theta$ -- in fact these densities are the product of Bayesian inference in the sense that finding out their properties and estimating them is one of the primary goals of Bayesian analyses.

The prior probability density of the model parameters $\pi(\theta)$ can also have a variety of shapes and natures but ideally it should be kept uninformative\footnote{Sometimes ``uninformative'' is used to denote a probability density that all scientists can agree on really being an uninformative one, i.e. representing ``maximum amount of ignorance''. However, it remains a subjective issue and we do not define what we mean by uninformative here other than that the obtained results depend more on the likelihood than the prior in practice.} w.r.t. the likelihood function in the sense that the likelihood is a narrower probability density (has e.g. greater Fischer information) in the parameter space. This means that the likelihood \emph{overwhelms} the prior, though sometimes it is necessary to reverse the situation by using informative priors that in turn overwhelm the likelihood -- especially if the likelihood function does not contain information on some aspects of the modelled system.

Priors are the only subjective part of Bayesian analyses -- the rest consists of computations an is largely neglected throughout this article with few exceptions, though sometimes the computational difficulties are considerable problems on their own. However, some assumptions have to be made regarding prior probability densities. This is evident because a flat, i.e. uniform, prior is still a prior density (one that all frequentists using likelihoods assume). Similarly, fixing parameters (e.g. fixing eccentricity such that $e=0$ in planet searches) corresponds to a delta-function prior of the form $\pi(e) = \delta(e)$ that has all the density at zero and none elsewhere. Furthermore, prior probabilities of different models do not have to be equal but some models might appear more realistical \emph{a priori} and it would then be desirable to assign them greater prior probabilities in practice. Finally, the collection of candidate models is also selected \emph{a priori} and because comparisons of these models is actually indistinguishable from comparisons of different prior models\footnote{For instance, comparison of two models containing $k$ and $k+1$ Keplerian signals is equivalent to comparing two priors that are otherwise the same, but the other one has a delta-function form for the amplitude of the $k+1$th signal such that $\pi(K_{k+1}) = \delta(K_{k+1})$ that peaks at zero.}, the choice of candidate models is of very high importance in Bayesian analyses of scientific data.

\section{Bayesian inference with dynamical information}\label{sec:dynamics}

In case of detections of exoplanets in systems of multiple planets, there is a very useful additional source of information -- the Newtonian (or post-Newtonian if necessary) laws of motion. Because we cannot expect to detect an unstable planetary system in practice -- for the very reason that because of instability it would not exist in the first place unless we were observing it exactly at the time when its chaotic nature causes collisions or ejections of the bodies in the system -- we can say that the prior probability of detecting something unstable is almost negligible.

In the Bayesian context, this additional information from dynamical analyses of such planetary systems can be written as
\begin{equation}\label{bayes_dynamical}
  \pi(\theta | m, \mathcal{S}) = \frac{l(m, \mathcal{S} | \theta) \pi(\theta)}{P(m, \mathcal{S})} = \frac{l(m | \mathcal{S}, \theta) l(\mathcal{S} | \theta) \pi(\theta)}{P(m, \mathcal{S})} ,
\end{equation}
where $\mathcal{S}$ denotes this ``dynamical information'' in terms of stability constraints of the system. Because Newton's laws, or more generally the laws of gravity, do not depend on the measurements we have possibly obtained but the results we obtain must depend on them (unless we have found something that violates them -- a significant discovery indeed), we have written the last equality in the above equation. In particular, the likelihood containing the dynamical information, $l(\mathcal{S} | \theta)$, needs to be defined in some suitable way to be able to use the Eq. (\ref{bayes_dynamical}).

\subsection{Analytical criteria for dynamical stability}

The dynamical likelihood in Eq. (\ref{bayes_dynamical}) can be written simply by using some simple stability criteria and by defining the likelihood in such a way that it is negligible for systems not satisfying these criteria while it has a non-negligible constant value for systems that do satisfy the criteria.

One simple such criterion is that the planets in a given system should not experience orbital crossings. Assuming co-planarity of the orbits, this leads to an approximate criterion for the eccentricities and semi-major axes ($a$) of the planets $i$ and $j$ such that $a_{i}(1+e_{i}) < a_{j}(1-e_{j})$ when the $i$th planet is the inner one and the $j$th is the outer one. This criterion can further be improved by requiring, for instance, that the planets stay out of each others Hill spheres at all times. However, as a very rough criterion, this expression can only be used as a starting point and cannot be expected to provide much additional constraints in practice unless the proposed planetary system is clearly unstable and one of some of the corresponding signals are actually caused by something else but planets.

A slightly more useful criterion is e.g. the approximated Lagrange stability criterion for two subsequent planets defined as \cite{barnes2006}
\begin{equation}\label{stability}
  \alpha^{-3} \bigg( \mu_{1} - \frac{\mu_{2}}{\delta^{2}} \bigg) \big( \mu_{1} \gamma_{1} + \mu_{2} \gamma_{2} \delta \big)^{2} > 1 + \mu_{1}\mu_{2} \bigg( \frac{3}{\alpha} \bigg)^{4/3} ,
\end{equation}
where $\mu_{i} = m_{i} M^{-1}$, $\alpha = \mu_{1} + \mu_{2}$, $\gamma_{i} = \sqrt{1 - e^{2}_{i}}$, $\delta = \sqrt{a_{2}/a_{1}}$, $M = m_{\star} + m_{1} + m_{2}$, $m_{i}$ is the planetary mass, and $m_{\star}$ is stellar mass. Again, we simply set $l(\mathcal{S} | \theta) = c$ if the criterion is satisfied and $l(\mathcal{S} | \theta) = 0$ otherwise (we call the set of stable orbits in the parameter space $B \subset \Theta$). This criterion can also be used to estimate briefly whether a given observed system is indeed stable or not \cite{tuomi2012,tuomi2013a,tuomi2013b}.

However, the above criteria do not take into account e.g. orbital resonances and are only very approximate criteria in practice. Therefore, better methods are needed to take into account the dynamical information in practice.

\subsection{Dynamical information from posterior samplings}

If the shape of the posterior density of the model parameters is already known and a sample has been drawn from it e.g. using one of the various posterior sampling algorithms \cite{metropolis1953,hastings1970,haario2001}, we can use this sample in assessing the shape of the dynamical likelihood function $l(\mathcal{S} | \theta)$. We assume that such a statistically representative sample has been drawn from the posterior density and that it exists in a form of a Markov chain such that $\theta_{i} \sim \pi(\theta | m), i=1, ..., K$. These $K$ parameter vectors can now be used as initial states of direct numerical integrations of planetary orbits.

As performing numerical integrations of planetary orbits is a simple task with suitable algorithms, such as the Bulirsch-Stoer algorithm \cite{bulirsch1966}, we assume that each $K$ initial states of the system have been integrated from $t_{0} =  0$ to $t_{N} = T$, where the duration of these integrations, $T$, is typically set to millions to hundred million years. Also, some crite5ria have been chosen to determine whether a system is stable or not (i.e. collisions, increasing $a_{i}$ without limit for some $i$, etc.). We divide the duration $T$ into $N$ intervals such that $t_{i} = iT/N$. With this notation, and with $\theta_{i}$ as an initial state of orbital integrations for all $i = i, ..., K$, we have $K$ chains of $N$ vectors that we denote as $\theta_{i}^{j}, j=1, ..., N$, where $\theta_{i}^{j} = \theta_{i}(t_{j})$. Each of these chains of vectors represent the orbital evolution of the system as a function of time.

Hence, we can approximate the posterior probability of finding $\theta \in I_{l} \subset \Theta$ of dynamical information and data for each $n$-interval $I_{l}$ as
\begin{eqnarray}\label{interval_approximation_dynamics}
  P(\theta \in I_{l} | \mathcal{S}, m) && \approx \frac{1}{K} \sum_{i=1}^{K} P(\theta \in I_{l} | \mathcal{S}, \theta_{i}) \nonumber\\
  && \approx \frac{1}{KN}\sum_{i=1}^{K} \sum_{j=1}^{N} \mathbf{1}_{l}(\theta_{i}^{j}) \mathbf{1}(\theta_{1}^{j}),
\end{eqnarray}
where
\begin{equation}\label{indicator2}
\mathbf{1}_{l}(\theta) =
  \left\{ \begin{array}{l}
  1 \textrm{ if } \theta \in I_{l} \\
  0 \textrm{ otherwise} 
  \end{array} \right. \textrm{ and }
\mathbf{1}(\theta) =
  \left\{ \begin{array}{l}
    1 \textrm{ if } \theta \in B \\
    0 \textrm{ otherwise}
  \end{array} \right. \nonumber
\end{equation}

In the above expression, the second indicator function denotes whether a given chain of vectors corresponds to an unstable configuration or not, i.e. whether any member of the chain is outside the set of stable orbits $B$. The first indicator function denotes the probability of finding the parameter vector in the given $n$-interval and weights each part of the set $B$ with this probability. Thus, this expression sets negligible weight to unstable configurations and assigns nonzero probabilities for the stable ones and can be interpreted as the desired posterior probability density of the model parameters given the measurements and the dynamical information in Eq. (\ref{bayes_dynamical}). Applying this equation is not difficult in practice given that one can increase $K$, $N$, and $T$ such that the dynamical analyses help in ruling out unstable areas of the parameter space and the results become statistically representative \cite{tuomi2013a}.

\section{Planet detection criteria}\label{sec:criteria}

When the goal is to find planetary signatures from noisy data, the criteria for detecting confidently such signatures are important in order to extract weak signals hidden in the noise from the data and to avoid detecting false positives or signals arising from other periodic or quasi-periodic sources such as the movement of the celestial bodies of the Solar System; stellar rotation coupled with magnetic activity, starspots, and other activity-induced phenomena of the stellar surface; of instrumental source as a consequence of lack of sufficient instrumental stability or problems in calibration.

The first part of these criteria can be summarised briefly as follows \cite{tuomi2012}. First, it is required that the posterior probability of a model with $k+1$ signals is significantly greater than that of a model with only $k$ signals. This can be expressed as $P(\mathcal{M}_{k} | m) > \alpha P(\mathcal{M}_{k-1} | m)$ for a selected threshold $\alpha > 1$. Several authors use $\alpha = 150$ that has been interpreted as corresponding to ``significant evidence'' \citep{kass1995,feroz2011,tuomi2012} but the exact choice of this threshold remains subjective and depends only on how confident one wants to be when making decisions based on model comparisons.

The second criterion is that, in case of radial velocities that will be used as examples below, the radial velocity amplitudes of all signals are statistically distinguishable from zero, i.e. their Bayesian credibility sets (BCSs)\footnote{The BCS, a subset of the parameter space $\Theta$, of threshold $\delta$ is defined for the posterior density $\pi(\theta | m)$ as $\mathcal{D}_{\delta} = \bigg\{ \mathcal{D}_{\delta} \subset \Theta : \int_{\theta \in \mathcal{D}_{\delta}} \pi(\theta | m) = \delta, \pi(\theta | m)|_{\theta \in \partial \mathcal{D}_{\delta}} = c \bigg\}$.} do not overlap with zero for a selected threshold $\delta \in [0,1]$. In practice this means that the signals are constrained from below in a sense that they cannot be said to be consistent with zero, which in turn would imply that they are not significantly present in the data. In practice there are several examples of signals that do not satiusfy this criterion or satisfy it only just and it should be taken into account when detecting periodic variations of planetary origin \cite{tuomi2012,tuomi2013a,tuomi2013b}.

In addition to having a well-constrained amplitude, the period of a signal also has to be well-constrained in the sense that it has clear lower and upper limits in the parameter space (with suitable credibility). If this was not the case, a signal could not be said to be periodic and could not be readily interpreted as being caused by the Doppler effect of planetary origin. However, this one, and the other two above criteria are only mathematical ones applicable to any detection problem of periodic signals. There are also other, physical, criteria that have to be satisfied to be able to claim that a given signal is of planetary origin. 

One of these physical criteria was already mentioned above: a planetary system has to be stable in long term. Therefore, the orbital parameters obtained from the data have to correspond to a dynamically viable system. Furthermore, the signals should not have counterparts in the stellar activity indicators, which would cast considerable doubt on their planetary nature. However, we do not discuss these criteria further here for brevity.

\section{Modelling radial velocities in practice}

Radial velocity method is, together with the transit photometry method, the most efficient method of finding planets around other stars. However, because it is not limited to inclinations that enable transits, it is much more efficient in finding planets around nearby stars \cite{anglada2012,anglada2012b,tuomi2012,tuomi2013a,tuomi2013b}. However, in case of low-mass planets in multiple systems, and in the precens of radial velocity noise of comparable magnitude to the planetary signals, the typical periodogram-based planet searches fail dramatically and prevent the detections of low-mass planet candidates \cite{lovis2011,anglada2012b}.

One of the reasons for this failure is that usually radial velocity data is binned by calculating e.g. the average of few velocities within an hour or within a given night or so\footnote{Typically this binning is performed in an unknown manner and its details, i.e. weighting and choice of estimates, is not reported in the literature reporting radial velocity data.}. As the process of averaging several measurements is clearly not a bijective mapping, it is then impossible to return from the binned data usually reported in the literature to the original set and investigate the validity of the binning algorithm. Moreover, binning will always result in loss of information because it essentially corresponds to an artificial decrease in the number of precious measurements. Therefore, it is hard to find any justification to binning other than it makes the resulting computations easier by making the obtained binned velocities closer to independent that makes the periodogram analyses more suitable. For these reasons, more robust approaches are needed.

Instead, modelling the noise as realistically as possible can lead to much more trustworthy results than would be possible using any binning algorithm \cite{tuomi2013b}. The first possibility would be to have the binning procedure as a part of the statistical model, which enables comparisons of different procedures in the context of Bayesian model comparisons. One such procedure effectively corresponding to binning is a noise model with moving average (MA) component \cite{tuomi2012c,tuomi2013a,tuomi2013b}. This statistical model can be written briefly as
\begin{equation}\label{moving_average}
  m_{i} = f_{k}(t_{i}) + \epsilon_{i} + \sum_{j=1}^{p} \phi_{j} \big[ m_{i-j} - f_{k}(t_{i-j}) \big] ,
\end{equation}
where measurement $m_{i}$ at epoch $t_{i}$ is modelled using the function $f_{k}$ and some convenient white noise component $\epsilon_{i}$ that can be e.g. a combination of instrumental white noise, as determined by the instrument pipeline producing the velocity measurements, and an additional source of noise, sometimes referred to as \emph{stellar jitter}.

The analyses of radial velocities from HARPS and other instruments indicate that this noise model is typically much better than pure white noise model \citep{tuomi2012c,tuomi2013a,tuomi2013b}. Furthermore, information is not lost if the MA coefficients $\phi_{i}$ are selected conveniently or used as free parameters of the statistical model.

\subsection{The $\tau$ Ceti velocities}

\begin{figure}
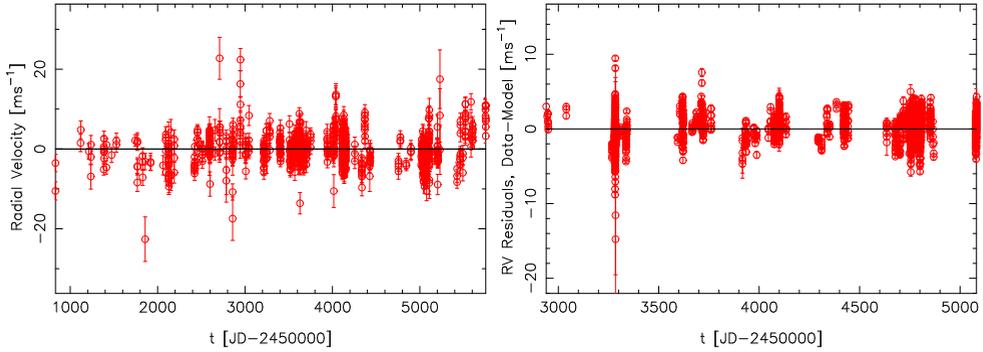

\centering
\includegraphics[angle=270,width=0.45\textwidth]{rvdist00_curvec_rv_HD10700e.ps}
\includegraphics[angle=270,width=0.45\textwidth]{rvdist00_residc_rv_HD10700.ps}
\caption{UCLES (left) and HARPS (right) radial velocities of $\tau$ Ceti.}
\label{fig:velocities}
\end{figure}

The nearby Solar-type star $\tau$ Ceti (HD 10700) has been the most frequent target star of radial velocity surveys in the past \cite{wittenmyer2006,pepe2011,tuomi2013b} because it is the closest single Solar-type star to our own system. The radial velocity curve of this star has been considered ``flat'' despite $\sim$ 4400 HARPS high-precision velocities, $\sim$ 1000 AAT/UCLES radial velocities, and $\sim$ 600 HIRES precision velocities from the 10m Keck telescope and no planets have been reported to orbit $\tau$ Ceti \cite{wittenmyer2006,pepe2011}. We show the UCLES and HARPS velocities of this star in Fig. \ref{fig:velocities} \cite{tuomi2013b}. These long time-series have been shown to contain significant amounts of covariance, or correlations between subsequent measurements in different timescales \cite{tuomi2013b}, and the MA model of Eq. (\ref{moving_average}), with exponential smoothing, has been used to model the velocities succesfully \cite{tuomi2013b}.

In particular, modelling the HARPS data in Fig. \ref{fig:velocities} (right panel) with the MA model decreased the amount of excess white noise in the data tremendously from 1.60 [1.51, 1.69] to 1.06 [1.02, 1.11] ms$^{-1}$ \cite{tuomi2013b} -- as denoted using the maximum \emph{a posteriori} (MAP) estimates and the corresponding BCSs with $\delta = 0.99$. As the correlations in the noise are more deterministic and can be removed according to the model, this decrease is very significant improvement in comparison to the traditional white noise models. Similar improvements have also been observed for the HIRES and UCLES data \cite{tuomi2013b}

Furthermore, such decrease in the noise, and the most importantly the ability to make the excess noise remaining in the data more white \cite{tuomi2013a,tuomi2013b}, i.e. independent, enables us to use even the more traditional methods in analysing the data, such as the Lomb-Scargle periodograms \cite{lomb1976,scargle1982}, to visualise the results. We compare such periodograms of the ``raw'' HARPS data shown in Fig. \ref{fig:velocities} (right panel) and of residuals after removing the correlations in the model according to the MAP estimates of an MA(10) model \cite{tuomi2013b}. These periodograms are shown in Fig. \ref{fig:periodograms}.

\begin{figure}
\centering
\includegraphics[angle=270,width=0.9\textwidth]{tauceti_raw.ps}
\includegraphics[angle=270,width=0.9\textwidth]{tauceti_0.ps}
\caption{Lomb-Scargle periodograms of the HARPS velocities before (top) and after (bottom) removing the noise correlations according to the MAP estimates of the MA(10) model. The dotted, dashed, and dot-dashed lines indicate the analytical 10\%, 1\%, and 0.1\% false alarm probabilities.}
\label{fig:periodograms}
\end{figure}

Clearly, the large number of peaks in excess of the analytical false alarm probabilities (FAPs) in Fig. \ref{fig:periodograms} (top panel) do not indicate the precense of a variety of significant periodicities in the data but that the assumptions underlying the periodograms are not satisfied. Instead, in the bottom panel of Fig. \ref{fig:periodograms}, the situation is visually more realistical and the significance of the peaks exceeding the 10\% FAP at periods of roughly 14, 35, 300, and 600 days should be investigated further \cite{tuomi2013b}.

\begin{figure}
\centering
\includegraphics[angle=270,width=0.9\textwidth]{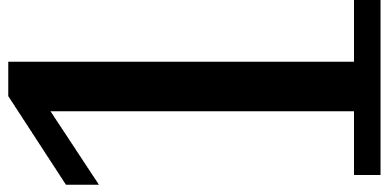}
\includegraphics[angle=270,width=0.9\textwidth]{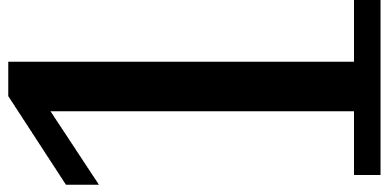}
\caption{As in Fig. \ref{fig:periodograms} but after removing correlations from the UCLES (top) and HIRES (bottom) data sets.}
\label{fig:periodograms2}
\end{figure}

After removing the correlations from the UCLES and HIRES data sets -- this time in the time-scale of $\sim$ 10 days -- the Lomb-Scargle periodograms of these sets appear flat and do not appear have any significant powers (Fig. \ref{fig:periodograms2}). However, there are indications that models taking into account noise correlations in a time-scale of $\sim$ 10 days can adapt to the signals and decrease their significances in the residual periodograms (at least this has been observed for HARPS data of GJ 581) \cite{baluev2012,tuomi2012c}.

\section{Analysis of HD 40307 radial velocities}\label{sec:hd40307}

\begin{figure}
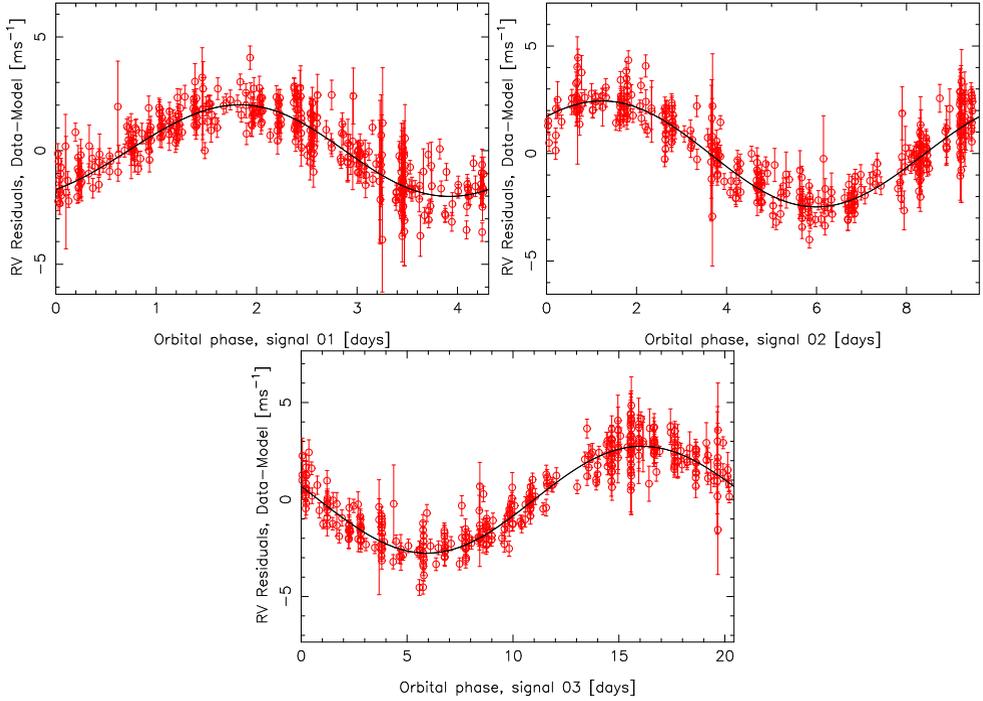

\centering
\includegraphics[angle=270, width=0.45\textwidth]{rvdist06_scresidc_rv_HD40307b_1.ps}
\includegraphics[angle=270, width=0.45\textwidth]{rvdist06_scresidc_rv_HD40307b_2.ps}

\includegraphics[angle=270, width=0.45\textwidth]{rvdist06_scresidc_rv_HD40307b_3.ps}
\caption{Phase-folded orbits of HD 40307 b (top left), c (top right), and d (bottom) together with the HARPS velocities.}
\label{fig:hd40307_orbits}
\end{figure}

HD 40307 is known to be a system of three super-Earths with orbital periods of 4.31, 9.62, and 20.44 days and minimum masses of 4.3, 7.0, and 10.5 M$_{\oplus}$, respectively \cite{mayor2009,tuomi2013a}. Obtaining the signals of these three super-Earths from HARPS-TERRA velocities \cite{anglada2012c,tuomi2013a} is rather easy and we have plotted the corresponding phase-folded signals in Fig. \ref{fig:hd40307_orbits}. However, the three planetary companions were reported based on analyses of binned HARPS-CCF velocities \cite{mayor2009}. Therefore, binning the data, and using a sub-optimal spectral reduction method, i.e. the cross-correlations function (CCF) method \cite{anglada2012c}, might disable a reliable detection of additional low-amplitude signals of planetary origin that might still be present in the data in addition to the three clear signatures of super-Earths.

Indeed, moving sequentially by subtracting the signals corresponding to the greates posterior probabilities (and strongest periodogram powers) from the data, and modelling the noise carefully with an MA(3) model enables the detections of three new signals in the HD 40307 velocities \cite{tuomi2013a}.

When looking at the residual periodograms of three-Keplerian model for the unbinned velocities (Fig. \ref{fig:periodograms3}, top panel), it can be seen that there are two powers exceeding the 10\% FAP in the data. However, without binning, and after actually throwing part of the spectral information away in the sense that the velocities are calculated only from the redmost part of the HARPS spectra likely containing less activity-induced variation \cite{tuomi2013a}, the significances of the powers increase (Fig. \ref{fig:periodograms3}) and indicate clearly that e.g. the 51-day peak is a significant periodicity in the data.

\begin{figure}
\centering
\includegraphics[angle=270, width=0.9\textwidth]{hd40307_bin_3.ps}
\includegraphics[angle=270, width=0.9\textwidth]{hd40307_3.ps}
\caption{As in Fig. \ref{fig:periodograms} but for the binned HARPS-TERRA velocities of HD 40307 (top) and for the unbinned ones after obtaining the velocities from the red part of the HARPS spectra only (bottom).}
\label{fig:periodograms3}
\end{figure}

\begin{figure}
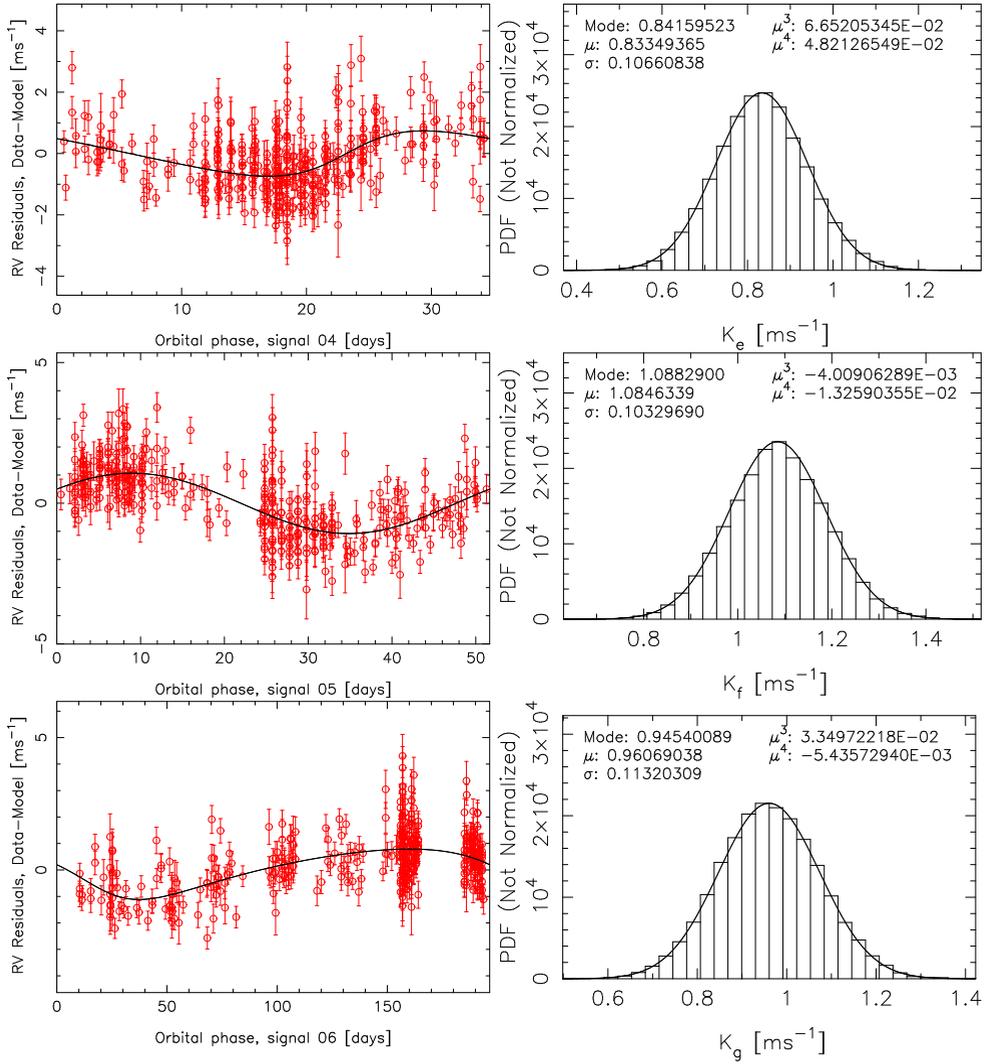

\centering
\includegraphics[angle=270, width=0.45\textwidth]{rvdist06_scresidc_rv_HD40307e_4.ps}
\includegraphics[angle=270, width=0.45\textwidth]{rvdist06_rv_HD40307e_dist_Ke.ps}

\includegraphics[angle=270, width=0.45\textwidth]{rvdist06_scresidc_rv_HD40307e_5.ps}
\includegraphics[angle=270, width=0.45\textwidth]{rvdist06_rv_HD40307e_dist_Kf.ps}

\includegraphics[angle=270, width=0.45\textwidth]{rvdist06_scresidc_rv_HD40307e_6.ps}
\includegraphics[angle=270, width=0.45\textwidth]{rvdist06_rv_HD40307e_dist_Kg.ps}
\caption{Phase-folded orbits and velocity amplitudes of the three new candidates around HD 40307.}
\label{fig:candidate_criteria}
\end{figure}

According to Bayesian analyses of the velocities of HD 40307, there are actually three additional periodicities in the data that satisfy all the detection criteria presented in Section \ref{sec:criteria}. This implies that there are actually six planets orbiting HD 40307 \cite{tuomi2013a}. We demonstrate this by showing the phase-folded signals of these candidates and the posterior probability densities of their radial velocity amplitudes in Fig. \ref{fig:candidate_criteria}. Interestingly, the outermost one of these planet candidates is in the liquid water habitable zone of the star \cite{selsis2007} and appears to be one of the best candidate for a habitable world out of those detected.

When analysing the HARPS-TERRA radial velocities of HD 40307, we did not use uninformative (in any practical meaning of uninformative) prior densities for the orbital eccentricities. Instead, we estimated that close-circular orbits of planets in any given system are more probable \emph{a priori} than more eccentric ones, and that extremely eccentric orbits, with e.g. $e > 0.6$, are very improbable in practice because they would certainly make such sysmtes of low-mass planets unstable due to orbital crossings. Furthermore, it is a well-known (but rather poorly understood) fact that analyses of radial velocity measurements containing signals with low eccentricities, tend to yield overestimates for the eccentricities \cite{zakamska2011}. For these reasons, we chose the prior such that $\pi(e) \propto \mathcal{N}(0,0.3^{2})$ \cite{tuomi2012,tuomi2013a}.

\begin{figure}
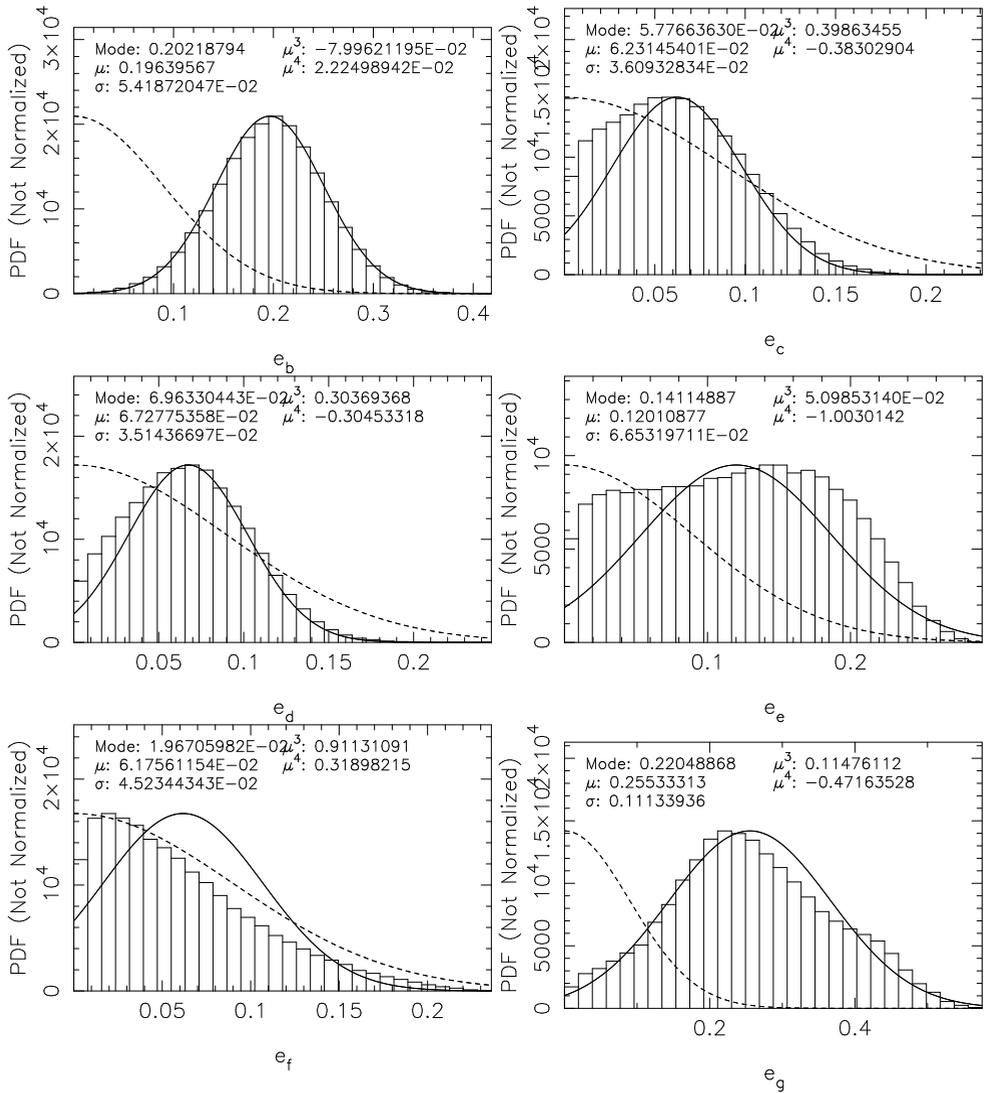

\centering
\includegraphics[angle=270, width=0.45\textwidth]{rvdist06_rv_HD40307e_dist_eb.ps}
\includegraphics[angle=270, width=0.45\textwidth]{rvdist06_rv_HD40307e_dist_ec.ps}

\includegraphics[angle=270, width=0.45\textwidth]{rvdist06_rv_HD40307e_dist_ed.ps}
\includegraphics[angle=270, width=0.45\textwidth]{rvdist06_rv_HD40307e_dist_ee.ps}

\includegraphics[angle=270, width=0.45\textwidth]{rvdist06_rv_HD40307e_dist_ef.ps}
\includegraphics[angle=270, width=0.45\textwidth]{rvdist06_rv_HD40307e_dist_eg.ps}
\caption{Estimated posterior densities of the eccentricities of the six planet candidates orbiting HD 40307 (histograms) and the corresponding prior densities of the eccentricities (dotted curve). The Solid curve represents a Gaussian density with the same mean and variance as the distribution.}
\label{fig:priors_posteriors}
\end{figure}

According to the comparison of posteriors and priors in Fig. \ref{fig:priors_posteriors}, the likelihood still overwhelms the priors -- at least, the prior does not appear to have a dominant role. This can be seen because eccentricities that are penalised rather heavily by the prior still have significant posterior values.

\begin{figure}
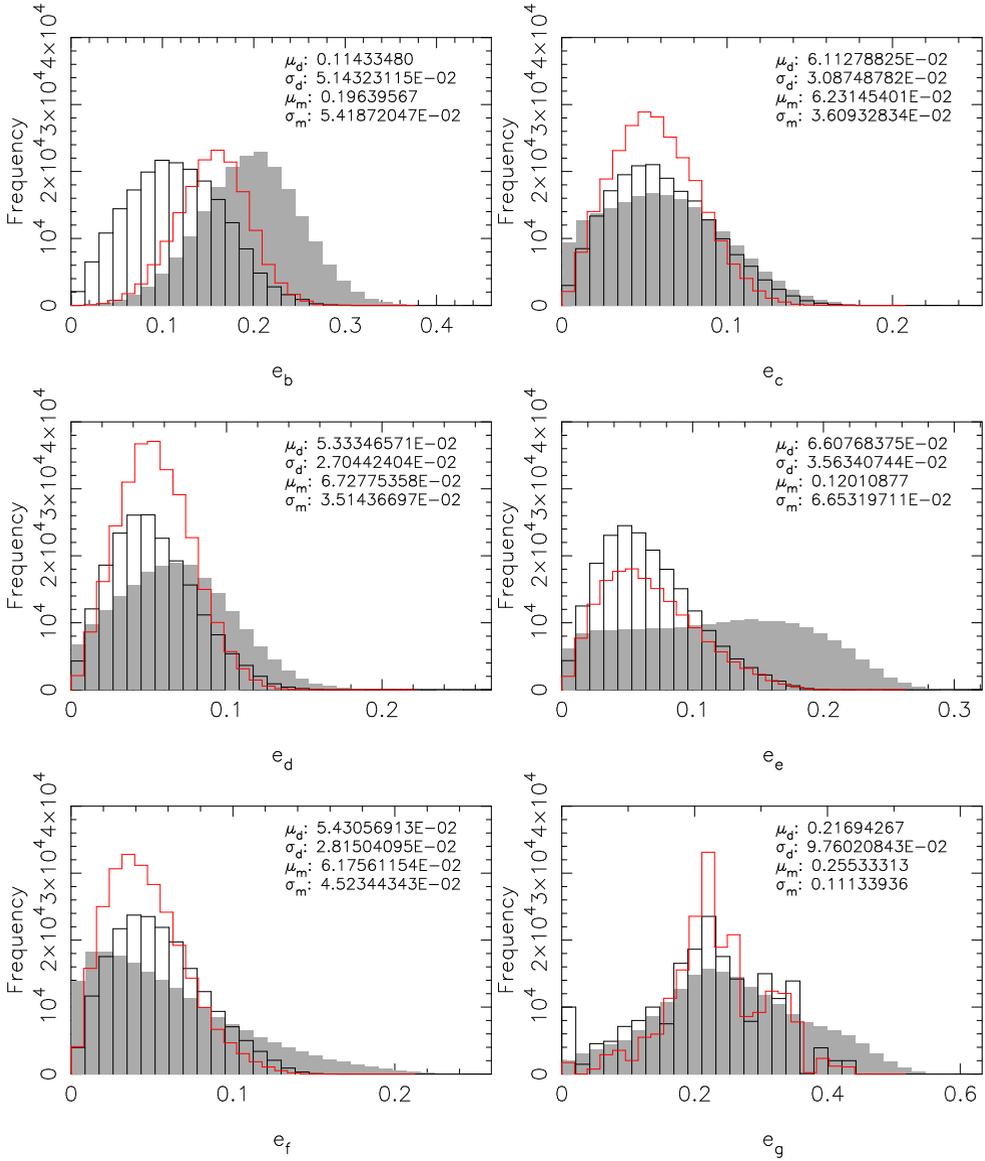

\centering
\includegraphics[angle=270, width=0.45\textwidth]{rvdist06_rv_HD40307e_nat_hist_eccs2_b.ps}
\includegraphics[angle=270, width=0.45\textwidth]{rvdist06_rv_HD40307e_nat_hist_eccs2_c.ps}

\includegraphics[angle=270, width=0.45\textwidth]{rvdist06_rv_HD40307e_nat_hist_eccs2_d.ps}
\includegraphics[angle=270, width=0.45\textwidth]{rvdist06_rv_HD40307e_nat_hist_eccs2_e.ps}

\includegraphics[angle=270, width=0.45\textwidth]{rvdist06_rv_HD40307e_nat_hist_eccs2_f.ps}
\includegraphics[angle=270, width=0.45\textwidth]{rvdist06_rv_HD40307e_nat_hist_eccs2_g.ps}
\caption{Posterior densities of the eccentricities (white histograms) of the six planet candidates orbiting HD 40307, dynamical likelihoods (grey histograms), and their combination (red curve).}
\label{fig:dynamical_updating}
\end{figure}

Furthermore, additional constraints for eccentricities can be readily obtained from dynamical analyses of the planetary orbits. Using the method described briefly in Eq. (\ref{interval_approximation_dynamics}), we plotted the resulting eccentricity posteriors, dynamical likelihoods, and their combinations in Fig. \ref{fig:dynamical_updating}. This corresponds to Bayesian updating of the posterior density from the data by the dynamical likelihood (grey histograms in Fig. \ref{fig:dynamical_updating}) and obtaining an updated posterior given both data and dynamical information (red curves in Fig. \ref{fig:dynamical_updating}).

It can be seen in Fig. \ref{fig:dynamical_updating} that the highest eccentricities of all the candidates in the HD 40307 system are basically penalised by the dynamical likelihood -- because they correspond to unstable orbital configurations. This example shows that information from dynamical analyses can be readily incorporated into Bayesian analyses of radial velocity data and can be used to better constrain the planetary orbits.

\section{Conclusions}

The Bayes' rule of conditional probabilities is an efficient and very practical way of inferring important information from noisy data. We have presented the typical formulations of Bayesian data analysis problems and showed some examples demonstrating the effectiveness of the combination of Bayes' rule and realistical noise models in practice.

In particular, when detecting periodic signals in noisy data, Bayesian formulation allows the definition of exact detection criteria that have to be satisfied to be able to conclude that a signal exists beyond reasonable doubt. Moreover, these criteria \cite{tuomi2012} can be expressed in a natural way and have intuitive probabilistic interpretations that leave few questions unanswered. When applying these criteria, the greatest caveat is the ability to assess whether the statistical models used to analyse the data, i.e. the likelihood and prior models, contain a sufficiently representative collection of models. If the detection criteria are satisfied for a variety of such models, there is little doubt about the existence of the signals. However, if this is the case for only one or few models, the physical origin of noise has to be understood better in order to make conclusions regarding the number of significant signals in the data.

The example of HD 40307 (Section \ref{sec:hd40307}) shows that a combination of priors, suitable likelihoods, and information from dynamical analyses can provide much more information on a given planetary system than any traditional periodogram-based methods. Furthermore, the combination of these sources of information is a natural application of the Bayes' rule and enables the combinations of any other sources of data in addition to these as well. Therefore, as the only logically consistent framework of data analysis \cite{berger1980}, the applications of Bayes' rule are plenty and it should be added to the basic toolbox of all astronomers in practice.

\end{document}